\documentclass[10pt, aps, prl, twocolumn, amsmath, amssymb, superscriptaddress, longbibliography]{revtex4-2}

\usepackage{graphicx}
\usepackage{dcolumn}
\usepackage{bm, amsfonts}
\usepackage{float}
\usepackage{braket}
\usepackage{xfrac}
\usepackage{xcolor, color, soul}
\usepackage{bbm}
\usepackage{upgreek}
\usepackage{newtxtext}
\usepackage{newtxmath}
\usepackage{setspace}
\usepackage[normalem]{ulem}

\usepackage[colorlinks=true , citecolor=black,urlcolor=blue]{hyperref}

\renewcommand{\section}[1]{\medskip\textbf{\textit{#1}} --}

\begin{document}

\title{Nonlinear collective dynamics and microwave comb generation in a diamond maser
}

\author{Christoph W. Zollitsch}
\affiliation{Department of Chemistry, Saarland University, Saarbr\"ucken, 66123, Germany}

\author{Jonas N. Bach}
\affiliation{Department of Chemistry, Saarland University, Saarbr\"ucken, 66123, Germany}

\author{Christopher W.M. Kay}
\affiliation{Department of Chemistry, Saarland University, Saarbr\"ucken, 66123, Germany}
\affiliation{London Centre for Nanotechnology, University College London, 17-19 Gordon Street, London, WC1H 0AH, UK.}

\author{Jonathan D. Breeze}
\affiliation{Department of Physics and Astronomy, University College London, Gower Street, London WC1E 6BT, UK}
\email{j.breeze@ucl.ac.uk}

\begin{abstract}
Optically pumped room-temperature masers are a promising platform for driven-dissipative spin–photon physics beyond steady-state emission. Here we observe a sequence of dynamical thresholds in a room-temperature nitrogen-vacancy diamond maser as the optical pump is increased above the conventional masing threshold. 
The first threshold yields narrow-line continuous-wave emission, while a second threshold produces a periodic train of microwave pulses whose repetition rate generates a frequency-comb spectrum. At higher pump power, a third threshold gives rise to pulses with an oscillatory, frequency-chirped decay. Time-resolved measurements link the comb directly to self-pulsing dynamics and Fourier analysis of individual bursts reveals a broad distribution of frequencies consistent with collective spin-resonator dynamics in an inhomogeneously broadened ensemble. These results establish the diamond maser as a room-temperature platform for nonlinear nonequilibrium light–matter dynamics and microwave-comb generation.
\end{abstract}

\maketitle
The realization of a room-temperature maser operating under ambient conditions, utilizing optically pumped nitrogen-vacancy (NV$^-$) centers in diamond \cite{Breeze2018}, presents the opportunity where masers could enable transformative technological developments similar to that of lasers. Without requiring cryogenics or ultra-high vacuum, routes to miniaturization are viable, enabling widespread application in modern technologies. Increasing experimental interest suggests a flourishing research field where efforts are made to bring diamond masers towards application in ultra-low noise signal amplification \cite{Day2024, Sherman2022}. Alternative maser platforms are being investigated that exploit periodically driven ensembles of atoms as Floquet maser amplifiers \cite{Jiang2022, Jiang2022a} for advanced quantum sensing, while a continued study of fundamental spin-spin interactions in diamond masers has potential to realize ultra-narrow linewidth through superradiant masing \cite{Kersten2026}.
Conventional masers are especially well established in the field of metrology, being one of the most stable frequency standards \cite{Vessot2005}. They offer a single fixed mode, whereas frequency combs are an alternative highly accurate frequency standard and are widely used for high precision spectroscopy \cite{Fortier2019}. Combs are mostly developed in optical frequency regimes, but microwave frequency combs are of growing interest due to their compatibility with modern communication technologies. Research on microwave frequency combs targets several platforms, such as magnonic \cite{Xu2023, Wang2024}, superconducting circuits \cite{Jeong2025, Greco2026}, semiconductors \cite{Xiang2023, Sun2024}, and recently in microwave-pumped ensembles of NV$^-$ spins by exploiting the nonlinearity at a bistable point of collective resonator-spin coupling \cite{Wang2026}. 
Here, we report the observation of nonlinear behavior in a diamond maser. Self-oscillation is achieved by optical laser pumping of NV spins above the threshold pump rate. Upon increasing the pump rate further we observe additional thresholds, where the maser emission spectrum changes from a single mode to a comb structure. We characterize each regime by spectrum analysis and time-resolved transient spectroscopy. A periodic train of microwave pulses is observed beyond the regime of continuous-wave masing; Fourier analysis reveals a frequency-chirp modulation of the pulses for highest pump rates. Utilizing frequency combs in masers combines two of currently most accurate frequency standard technologies and holds potential for high precision metrology and sensing.

Maser oscillation is typically demonstrated by population inverted ensembles of two-level systems resonantly coupled to resonant modes of high quality factor resonators.
The inverted populations are maintained by employing multi-level pumping schemes. 
The pump threshold rate is the minimum pump rate necessary to achieve level inversion for continuous emission of microwaves, typically achieved through an optical pump $w_\mathrm{opt}$ with green laser light. For NV-centers in diamond, the maser transition is formed by splitting the electronic $\ket{0}$ and $\ket{-1}$ spin states in the triplet ground state manifold by a magnetic field aligned along an NV defect axis \cite{Breeze2018}. Although the pumping process encompasses a total of eight energy levels, it can be mapped onto an effective two-level system \cite{Zollitsch2025}, with a scaled pump rate $w = \eta w_\mathrm{opt}$. 
For pump rates larger than the masing pump threshold $w_\mathrm{thr}$ ($w_\mathrm{opt} > w_\mathrm{thr}$) losses in the resonator are overcome by emission from the gain medium resulting in maser self-oscillation.  However, we have found that this stable maser phase is bound by additional thresholds as pump rates increase. Here, we report the observation of two additional thresholds beyond the primary masing threshold $w_\mathrm{thr}$ where emission undergoes distinct spectral changes, indicating that maser oscillators exhibit nonlinear dynamical behavior, which we characterize by steady-state and transient microwave spectroscopy.
\begin{figure}[t]
\begin{center}
\includegraphics{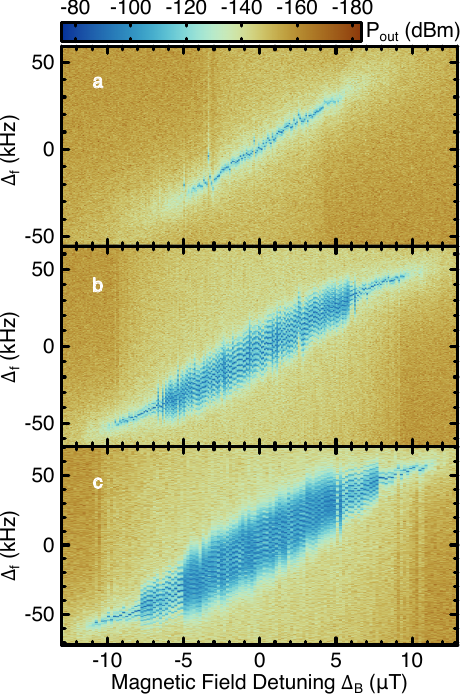}
\caption{
Logarithmic maser output power $P_\mathrm{out}$ as a function of magnetic field $\Delta_\mathrm{B}$ and frequency $\Delta_\mathrm{f}$ detuning for pump rates \textbf{a} $w =$ 6\,s$^{-1}$, \textbf{b} $w =$ 7.8\,s$^{-1}$ and \textbf{c} $w =$ 12.5\,s$^{-1}$. The resonator loaded $Q$-factor for all three spectra is 42,772. 
\label{fig:spec}
}
\end{center}
\end{figure}

The maser consists of an ensemble of NV$^-$ spin-defects within a synthetic diamond host with natural carbon abundance. The total NV concentration is $\approx$0.34\,ppm with equal distributions of the four possible $\left<{111}\right>$ defect axis orientations in the diamond structure. We tune the maser transition using the Zeeman effect with a magnetic field $B_0$ aligned to one $\left<{111}\right>$ defect axis. For fields larger than the NV zero-field splitting ($B_0 > 102.5$ mT), the spin states are pure and the $\ket{-1}$ state is energetically lower than the $\ket{0}$ state \cite{Tetienne2012, Patel2024}. Due to crystallographic orientation and a hyperfine splitting to $^{14}$N nuclear spins the absolute number of spins contributing to masing is $N$ = 4 $\times$ 10$^{13}$ \cite{Breeze2018}.
Spin-lattice and spin dephasing times have previously been characterized by pulsed ESR spectroscopy yielding $T_1 =$ 4.8\,ms and $T_2^* =$ 500\,ns, respectively. 
The diamond 
is placed inside a sapphire ring resonator operating at a resonance frequency $\omega_\mathrm{c}/2\pi = $ 9.23\,GHz. The sapphire ring is mounted inside a copper cavity to suppress radiation losses and provides an adjustable coupling between resonator and readout circuitry via a waveguide iris (see setup details in Ref. \cite{Zollitsch2023}). For minimal external coupling, the loss rate is $\kappa /2\pi =$ 108.9\,kHz. This corresponds to a loaded quality factor $Q$ of 42,772 where the internal Q-factor $Q_\mathrm{i}$ is determined to be 49,888. This is a significant enhancement over our previous setup \cite{Zollitsch2023}, whereby the sapphire ring holders were made of 3D-printed cyclic-olefin-copolymer (COC), instead of machined Polytetrafluoroethylene (PTFE) (see SI for details). Finally, the spin-photon coupling is assumed to be homogeneous over the sample dimension and is calculated to be $g/2\pi =$ 0.11\,Hz \cite{Zollitsch2023}. 

To generate maser oscillation, we optically pump the NV$^-$ spins with a 520\,nm multimode laser diode, while detecting the maser output power $P_\mathrm{out}$ with a spectrum analyzer. Figure \ref{fig:spec}\,(a,b,c) compares $P_\mathrm{out}$ as a function of magnetic field $\Delta_\mathrm{B}$ and detection frequency $\Delta_\mathrm{f}$ detuning, under continuous pump rates of 6\,s$^{-1}$, 7.8\,s$^{-1}$ and 12.5\,s$^{-1}$. The frequency detunings for each measurement are referenced to the respective resonance field of the central spin transition and the resonator frequency. Note that the absolute resonance conditions depend on the pump rate, due heating effects incurred by the laser and the temperature dependence of the resonator and zero-field splitting of the NV$^-$ centers \cite{Zollitsch2023}. The full maser emission spectra show three separate lines corresponding to the NV$^-$ spin transition $\ket{0} \mapsto \ket{-1}$ split by hyperfine interaction, where we only focus on the central transition in Fig.\,\ref{fig:spec}. For lower pump rates, the maser emission line is spectrally narrow, while at higher pump rates the lineshape changes significantly. In Figure \ref{fig:spec}\,(b) the emission changes to a broad frequency comb structure, containing several narrow lineshapes. At the edges beyond $\Delta_\mathrm{B} \pm$ 6\,$\mu$T, a narrow single mode is observed with an abrupt transition between the two regimes. At the highest pump rate, the emission spectrum exhibits an additional third regime within the interval $\left| \Delta_\mathrm{B} \right| <$ 5\,$\mu$T (see Fig. \ref{fig:spec}\,(c)). Here, the pronounced frequency comb structure in Fig. \ref{fig:spec}\,(b) is superimposed by a broad flat-topped lineshape. Observing the three different regimes in a single field-swept measurement is possible due to the quadratic dependence of $w_\mathrm{thr}$ on the resonator-spin detuning parameter $\Delta$ \cite{Zollitsch2025}. In the remainder, we will focus on zero magnetic field detuning ($\Delta_\mathrm{B} =$ 0\,T).

\begin{figure}[h]
\begin{center}
\includegraphics{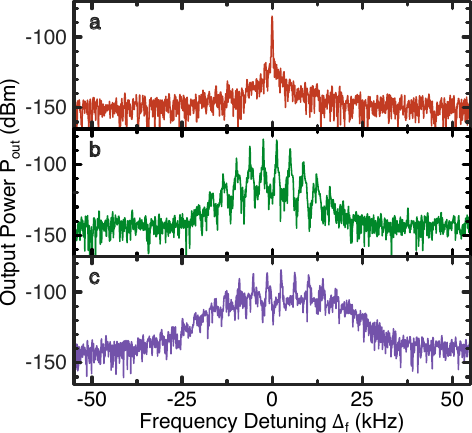}
\caption{
Logarithmic maser output power $P_\mathrm{out}$ as a function of frequency detuning $\Delta_\mathrm{f}$ at $\Delta_\mathrm{B} = 0$ (see Fig. \ref{fig:spec}) for pump rates \textbf{a} $w =$ 6\,s$^{-1}$, \textbf{b} $w =$ 7.8\,s$^{-1}$ and \textbf{c} $w =$ 12.5\,s$^{-1}$.
\label{fig:spec_cuts}
}
\end{center}
\end{figure}

Figure \ref{fig:spec_cuts} presents the three threshold regimes in more detail, showing $P_\mathrm{out}$ as a function of frequency detuning at $\Delta_\mathrm{B} =$ 0\,T and for $w =$ 6\,s$^{-1}$, 7.8\,s$^{-1}$ and 12.5\,s$^{-1}$ (see Fig. \ref{fig:spec_cuts}\,(a), (b) and (c), respectively). In the linear regime, the maser oscillation exhibits a narrow single-frequency lineshape (Fig.\,\ref{fig:spec_cuts} (a)). The lineshape is best described by a Voigt profile (convolution of Gaussian and Lorentzian), able to describe the fast dropping wings and narrow peak width of the emission signal. This yields a Gaussian standard deviation $\sigma_\mathrm{1st} =$ 69\,Hz $\pm$ 2\,Hz and a Lorentzian full-width at half-maximum $\Gamma_\mathrm{1st}$ = 51\,Hz $\pm$ 2\,Hz. 
The second threshold regime exhibits a frequency comb up to 10 teeth with a constant spacing of $\sim$\,3.8\,kHz (Fig. \ref{fig:spec_cuts}\,(b). Individual peaks are of comparable lineshape and width with $\sigma_\mathrm{2nd} =$ 50\,Hz $\pm$ 11\,Hz and $\Gamma_\mathrm{2nd}$ = 55\,Hz $\pm$ 12\,Hz. The larger error margins reflect a finite linewidth variation between individual comb teeth.
Figure \ref{fig:spec_cuts}\,(c) plots $P_\mathrm{out}$ in the third threshold regime. The frequency comb is buried in a broad emission spectrum, where only the tips of 7 comb-teeth are distinguishable. The tooth spacing of 3.8\,kHz and amplitudes remain unchanged to the second threshold regime and the linewidth $\sigma_\mathrm{3rd} =$ 89\,Hz $\pm$ 44\,Hz and $\Gamma_\mathrm{3rd}$ = 51\,Hz $\pm$ 66\,Hz. Large error margins result from the reduced signal-to-noise due to the broad emission background. 

\begin{figure}[h] 
\begin{center}
\includegraphics{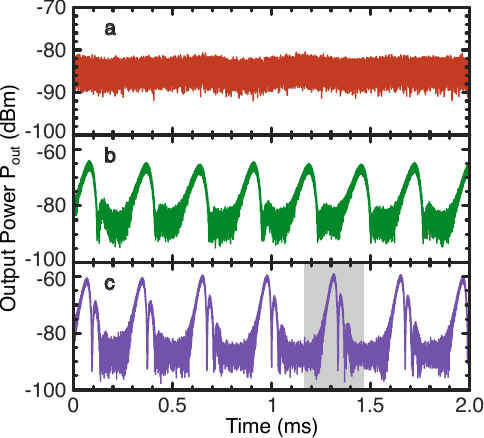}
\caption{
Transient logarithmic output power $P_\mathrm{out}$ as a function of time for the \textbf{a} first ($w =$ 6\,s$^{-1}$), \textbf{b} second ($w =$ 7.8\,s$^{-1}$) and \textbf{c} third ($w =$ 12.5\,s$^{-1}$) threshold regime at $\Delta_\mathrm{B} = 0$. A moving window average with an interval of \textbf{a} 80\,ns \textbf{b} 40\,ns \textbf{c} 200\,ns is applied.
\label{fig:trans_cuts}
}
\end{center}
\end{figure}

To gain more insight into the dynamics of the additional masing threshold regimes, we demodulate the emission signal and detect the in-phase and out-of-phase (IQ) quadratures. To align with steady-state experiments, we convert the quadratures into a magnitude signal, corresponding to the output power $P_\mathrm{out}$. The resulting transients are shown in Fig. \ref{fig:trans_cuts}\,(a)-(c) recorded at $\Delta_\mathrm{B} =$ 0\,T. The first regime exhibits continuous power emission (see Fig. \ref{fig:trans_cuts}\,(a)) at power levels of about -86\,dBm, matching the steady-state measurements. This constitutes a stable microwave emission within a single narrow frequency spectrum. In the second threshold regime, a drastic change occurs where the continuous emission changes to a periodic train of pulses (Fig. \ref{fig:trans_cuts}\,(b)). The pulse period $T =$ 261\,$\mu$s, matches the frequency comb-teeth spacing with $1/T =$ 3.8\,kHz, shown in Fig. \ref{fig:spec_cuts}\,(b). Fourier transforming a periodic train of pulses yields the observed frequency comb spectrum (see SI) and verifies its origin in nonlinear dynamics rather than being a result of continuous amplitude modulation of one system parameter \cite{Udem2002, Cao2014}. Each pulse peaks at -67\,dBm, a 100-fold increase in output power with respect to the continuous emission of the first threshold regime, but averages to comparable powers in the steady-state. Pulses last about 100\,$\mu$s and carry an integrated pulse energy of 16\,pJ. The shape exhibits an asymmetry in which the rise time is slower than the fall time. We attribute the asymmetry to the onset of superradiant emission \cite{Raimond1982, Rose2017, Angerer2018, Kersten2026}. Relaxation can be described by individual spin flips within a fully inverted coupled resonator-spin system \cite{Dicke1954}. 
As relaxation progresses, the spins experience a collective synchronization, leading to an enhancement of relaxation and thus a faster decay.
This effect becomes more pronounced in the third threshold regime, shown in Fig. \ref{fig:trans_cuts}\,(c). The emission still consists of a periodic train of pulses, but now exhibit an oscillatory decay. Peak emission powers increase to -60\,dBm and pulse duration increases to 150\,$\mu$s, prolonged by the oscillations. The integrated power of a single pulse doubles to 33\,pJ in comparison to the second threshold regime. Pulse periodicity exhibits an uncertainty, ranging from about 270\,$\mu$s to 300\,$\mu$s and matches the observed comb-teeth spacings between 3.3\,kHz and 3.7\,kHz.
The oscillatory decay of pulses is a characteristic feature of superradiance and has been reported in several experiments \cite{Skribanowitz1973, Kaluzny1983, Kersten2026}. In our recent work \cite{Zollitsch2025}, we theoretically investigated the composition of microwave emission from a diamond maser and found, for our parameter range, similar contributions from stimulated and superradiant emission. The theoretical prediction corroborates the identification of these spectroscopic features in terms of superradiance. This effect is enhanced by the increased optical pump rate. The maser emission into the resonator occurs at rates, which overcome resonator losses and enables partial reabsorption of excitation by the spins through the collective coupling. Together with the continuous optical pump, additional regions of the inhomogeneously broadened spin distribution exceed the masing threshold, triggering another emission burst. 

\begin{figure}[h]
\begin{center}
\includegraphics{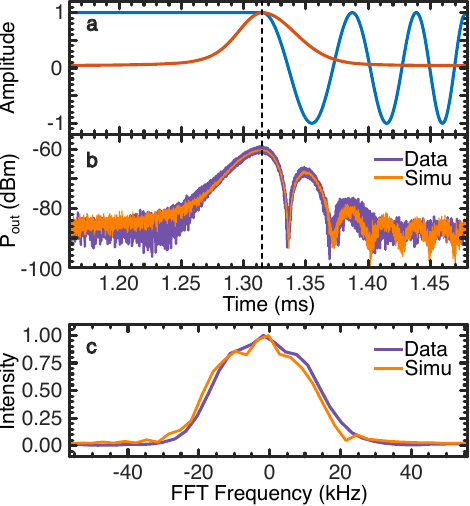}
\caption{
\textbf{a} Asymmetric Voigt profile (red) determined by piecewise fitting the pulse envelope in \textbf{b} and chirped pulse transient (blue), starting at the emission pulse maximum and sweeping frequency from 11.2\,kHz to 27.4\,kHz. \textbf{b} Single maser emission pulse in the third threshold regime (purple), as highlighted in Fig.\,\ref{fig:trans_cuts}\,(c). Orange solid line represents the combination of Voigt profile and chirped pulse of \textbf{a} with addition of Gaussian white noise of -90\,dBm. \textbf{c} Normalized fast Fourier transform of emission pulses in \textbf{b}, referenced to the intermediate frequency.
\label{fig:FFT}
}
\end{center}
\end{figure}

For a more detailed analysis of the oscillatory decay, we perform fast Fourier transforms (FFT) of individual pulses (see End Matter for details). Figure \ref{fig:FFT}\,(b) focuses on a single pulse of Fig. \ref{fig:trans_cuts}\,(c), indicated by a gray background.  
Figure \ref{fig:FFT}\,(c) shows the FFT frequency spectrum, normalized and referenced to a 2\,MHz intermediate frequency (IF) carrier. At its center, the IF carrier peaks and features additional shoulders symmetric about the center. Instead of a discrete frequency, the oscillatory decay exhibits a distribution of frequencies. This is a feature of chirped pulses, during which both amplitude and frequency change. To quantify the chirp behavior, we developed a model consisting of an asymmetric peak function modulated by chirped oscillations. In order to determine the underlying peak function, we fit the pulse envelope. Agreement is achieved with an asymmetric Voigt profile, in line with our analysis of the spectral emission peaks. 
Figure \ref{fig:FFT}\,(a) shows the envelope and is defined by a set of Lorentzian and Gaussian linewidth parameters for the rising and falling part of the pulse. Lorentzian FWHM $\kappa_\mathrm{rise} =$ 11 $\pm$ 0.4\,$\mu$s, $\kappa_\mathrm{fall} =$ 6.6 $\pm$ 0.5\,$\mu$s and Gaussian standard deviation $\sigma_\mathrm{rise} =$ 15.1 $\pm$ 0.3\,$\mu$s, $\sigma_\mathrm{fall} =$ 24\ $\pm$ 0.3\,$\mu$s. In a consecutive step, the envelope is modulated by a chirped oscillation, starting at its maximum (blue line in Fig. \ref{fig:FFT}\,(a), indicated by vertical dashed line). The chirp range is chosen to match the experimental data, achieved by a linear frequency sweep from 11.2\,kHz to 22.4\,kHz over a time interval of 165\,$\mu$s. Gaussian white noise with power spectral density of -90\,dBm is added to model the noise floor and create comparable conditions for FFT. The combined model agrees well with the experimental data, and its FFT reproduces the experimental frequency spectrum (orange line in Fig. \ref{fig:FFT}\,(b) and (c)). 

The FFT frequency spectrum shows a frequency distribution over a total range of about 40\,kHz. This closely resembles the broad emission spectrum observed in Fig. \ref{fig:spec_cuts}\,(c), suggesting the origin of this signal is due to the chirped oscillatory decay of the individual pulses. We attribute the frequency-chirp to be facilitated by the inhomogeneously broadened spin ensemble. During the oscillatory decay, 
the excitations will swap between the two systems as Rabi oscillations. However, in this case, the Rabi frequency is expected to be constant under the condition that the oscillating magnetic field of the resonator is homogeneous across the sample volume \cite{Angerer2016}, a condition fulfilled by our microwave resonator \cite{Breeze2018, Zollitsch2023}. Consequently, the change in frequency is induced by the spin ensemble. The spectral linewidth of the emission peaks $\Gamma$ and $\sigma$ are orders of magnitude narrower than the inhomogeneous linewidth $\gamma_\perp$, indicating that not all spins contribute to the pulse. The initial emission burst is derived from a small region of the inverted spin distribution, leaving a spectral hole. While the excitation resides inside the resonator, the optical pump continues to increase the inversion of the entire distribution. After the first Rabi cycle, part of the excitation is reabsorbed by the spins, partly restoring the inversion. This lifts off-center sections of the spin distribution beyond the masing threshold, adding to the emission of energy with an additional frequency detuning. Further Rabi cycles will repeat this effect, eventually producing emissions from more detuned sections of the distribution, increasing the oscillation frequency. The maximum number of oscillations depends on the pump rate $w$ and is bounded by the cooperativity $C$, representing the loss of coherence and energy of the hybrid system.


In conclusion, we explored maser dynamics beyond the stable single mode regime. Two additional distinct regimes of masing are observed for increased optical pump rates, where the continuous microwave emission changes to a periodic pulse pattern. The train of periodic pulses leads to the generation of a comb in the frequency domain, with up to 10 teeth with spacings of $\sim\,$3.8\,kHz . The pulsed emission exhibits further subdivision. Initially, the pulse trains consist of single superradiant bursts, which transition to pulses with an oscillatory decay at the highest pump rates. A Fourier analysis of individual pulses reveals that the oscillating decay is frequency chirped. 
We attribute this behavior to the inhomogeneously broadened NV$^-$ spin ensemble. While the collective resonator-spin ensemble coupling periodically exchanges excitations between the two systems, the continuous optical pump enables additional off-center sections of the spin distribution for masing. This leads to an enhancement of the oscillation frequency after each Rabi cycle.

Frequency comb generation through periodic pulsed emission patterns is often rooted in nonlinear dynamics \cite{Udem2002, Cao2014}. Although we experimentally observe nonlinear dynamics, we still need to identify the underlying system parameters. Comb teeth spacing and the maximal teeth number show no clear dependence on optical pump rate and require further study. In lasers, nonlinearity studies \cite{Abraham1985, Kazakov2017} demonstrate instabilities that result in complex transient behavior. The equations of motion for lasers can be rearranged to resemble the Lorenz equations \cite{Haken1975}, a model to describe nonlinear dynamics.
Expanding our theoretical framework on maser dynamics \cite{Zollitsch2025} to include time evolution will allow studying the underlying nonlinear mechanism and hence related to models developed for lasers. Higher optical pump rates would allow an advancement deeper into the nonlinear regime to further investigate the formation of chirped pulses. This is currently limited by the diamond heat management. Laser heating reduces $T_1$ and diminishes the effect of optical pumping \cite{Zollitsch2023} and prevents access to additional regimes of masing.
Understanding the generation of frequency combs in diamond masers will elevate their potential for applications in metrology. Typically, frequency combs require high pump powers, whereas the diamond maser can be operated with moderate to low laser powers, considerably easing application challenges.

\section{Acknowledgments}
We gratefully acknowledge funding from the Royal Society (URF/R1/191297), the UK Engineering and Physical Sciences Research Council (EP/S000798/2), the Deutsche Forschungsgemeinschaft (DFG, German Research Foundation) under project 550083266, and Saarland University.
\newline

\section{End Matter} 

\subsection{Fast Fourier Transform Procedure}
The FFT is performed on the complex quadrature components and is converted to a magnitude to relate to the other presented data. Bounded integration intervals during the FFT can cause artifacts close to zero frequency. To avoid any superpositions with low frequency FFT artifacts, we demodulate the experimental transient signal to an intermediate frequency (IF) carrier of 2\,MHz. The resulting frequency spectrum is offset by the IF and far detuned from any interfering artifacts.

\subsection{Microwave Spectroscopy}
The detection of the steady-state microwave emission from the maser is performed with a thinkRF R5500 Real-Time spectrum analyzer. For transient microwave spectroscopy we use the spectrum analyzer for demodulation, detuning the internal LO frequency by a finite IF frequency, for heterodyne detection with a Rigol DS8034-R oscilloscope. 

\subsection{Data Availability}

The data that supports the findings of this study is available upon reasonable request. 
\newline

\bibliography{Nonlinear_Maser.bib}

\end{document}


\title{Supplementary Material: Nonlinear collective dynamics and microwave comb generation in a diamond maser}

\author{Christoph W. Zollitsch}
\affiliation{Department of Chemistry, Saarland University, Saarbr\"ucken, 66123, Germany}
\email{christoph.zollitsch@uni-saarland.de}

\author{Jonas N. Bach}
\affiliation{Department of Chemistry, Saarland University, Saarbr\"ucken, 66123, Germany}

\author{Christopher W.M. Kay}
\affiliation{Department of Chemistry, Saarland University, Saarbr\"ucken, 66123, Germany}
\affiliation{London Centre for Nanotechnology, University College London, 17-19 Gordon Street, London, WC1H 0AH, UK.}

\author{Jonathan D. Breeze}
\affiliation{Department of Physics and Astronomy, University College London, Gower Street, London WC1E 6BT, UK}
\email{j.breeze@ucl.ac.uk}

\date{\today}

\maketitle

\section*{Supplementary Note 1: Fabrication of low microwave-loss dielectric holders}
 
Polytetrafluoroethylene (PTFE / Teflon) is commonly used as a holder material in low-loss microwave resonators, because of its low relative permittivity ($\varepsilon_r \approx 2.04$) and low loss tangent ($\tan\delta \sim 10^{-4}$) at 10\,GHz \cite{Riddle2003}. Our previous experimental setups used high-precision machined PTFE holders to hold the sapphire dielectric ring resonator in the center of the copper cavity. To minimize dielectric losses in the PTFE, the holders are designed to have minimal absolute weight. This came with a loss of stability, as PTFE is soft and can easily be deformed. Overall, the mechanical material properties set a limit in size and mass reduction, achieved by CNC machining. In recent years, 3d printing developed into a fast, precise, and cost effective alternative to mechanical machining. However, PTFE cannot be processed by conventional fused deposition modeling (FDM), so an alternative thermoplastic with good dielectric properties was needed. Among the materials available for FDM, cyclic olefin copolymer (COC) and polypropylene (PP) offer the best microwave performance, with loss tangents close to those of PTFE \cite{Stevens2022}. Split-post dielectric resonator measurements on FDM-printed COC samples yielded $\varepsilon_r \approx 2.24$ and $\tan\delta \approx 2.2 \times 10^{-4}$ at 1.1\,GHz \cite{Stevens2022}. This is more than an order of magnitude lower than the loss tangent of PLA ($9 \times 10^{-3}$ at GHz frequencies \cite{Lee2019}), a common thermoplastic in 3d printing. We chose COC over PP because PP tends to warp severely during printing \cite{Stevens2022}.
 
The holders were printed with COC filament (Creamelt) on a Bambu Lab X1 Carbon printer with an enclosed build chamber, using a 0.2\,mm stainless steel nozzle and a textured Polyetherimide build plate (all from Bambu Lab). Slicing was done in Bambu Studio (firmware 01.08.02.00; Bambu Studio 1.10.1.50). The nozzle temperature was 245\,$^\circ$C, the build plate was held at 70\,$^\circ$C, and the print speed was 50\,mm/s. We used a layer height of 0.08\,mm, 100\,$\%$ infill, and two perimeter walls. Automatic bed leveling and flow calibration were enabled for every print. Tree-type supports were added in the slicer for more complex geometries, and a 10\,mm brim with a 0.1\,mm gap was used to improve adhesion to the build plate.

\section*{Supplementary Note 2: Fourier Transform of Pulsed Maser Emission}

\begin{figure}[h]
\begin{center}
\includegraphics{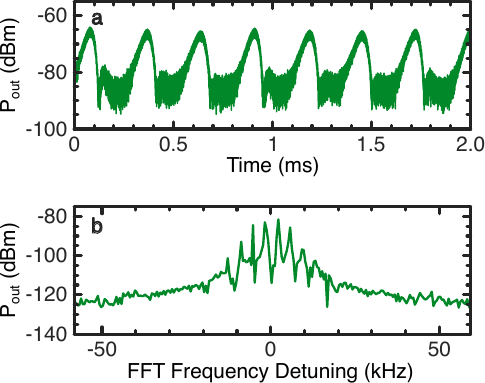}
\caption{
\textbf{FFT of transient maser output power.} 
\textbf{a} Transient logarithmic output power $P_\mathrm{out}$ as a function of time for the second ($w =$ 7.8\,s$^{-1}$) threshold regime at $\Delta_\mathrm{B} = 0$. A moving window average with an interval of 40\,ns is applied. \textbf{b} FFT of \textbf{a}, referenced to the intermediate frequency.
\label{figSI:trans_FFT}
}
\end{center}
\end{figure}

We observe additional masing thresholds beyond the initial, which enables continuous maser emission. For higher pump rates $w$ we identify two more thresholds where the emission profile drastically changes. The single narrow mode of the first regime changes into a frequency comb structure. Time dependent transient measurements show periodic pulsed maser emission, as discussed in detail in the main text. For an additional verification that that the periodic train of pulses yields a frequency comb spectrum, we fast Fourier transform (FFT) the transient maser output power $P_\mathrm{out}$ (see Fig. \ref{figSI:trans_FFT}\,(a)). The result is shown in Fig. \ref{figSI:trans_FFT}\,(b) and resembles a frequency comb, similar to Fig. 2\,(b) in the main text. The FFT is performed on the in-phase and out-of-phase quadrature signal and the result is converted to a logarithmic magnitude. The signal was demodulated to an intermediate frequency (IF), to which the FFT spectrum is referenced. The transient data has a time resolution of 8\,ns and an overall number of points of 250,000. Together, this yields in a 500\,kHz frequency resolution of the FFT.

\section*{Supplementary References}
\bibliography{Nonlinear_Maser.bib}